\documentclass[manuscript]{aastex}

\newcommand{\vden}{cm$^{-3}$}
\newcommand{\msun}{M$_{\sun}$}
\newcommand{\kms}{km~s$^{-1}$}

\newcommand{\cena}{Cen\,A}
\newcommand{\hto}{H$_{2}$O}

\shorttitle{Nuclear Water Maser Emission in Centaurus\,A}
\shortauthors{Ott et al.}

\begin{document}


\title{Discovery of Nuclear Water Maser Emission in Centaurus\,A}


\author{J\"urgen Ott}
\affil{National Radio Astronomy Observatory, P.O. Box O, 1003
  Lopezville Road, Socorro, NM 87801, USA}
\email{jott@nrao.edu}

\author{David S. Meier}
\affil{New Mexico Institute of Mining and Technology, 801 Leroy Place,
  Socorro, NM 87801, USA}
\affil{National Radio Astronomy Observatory, P.O. Box O, 1003
  Lopezville Road, Socorro, NM 87801, USA}
\email{dmeier@nmt.edu}

\author{Mark McCoy}
\affil{New Mexico Institute of Mining and Technology, 801 Leroy Place,
  Socorro, NM 87801, USA}
\email{mmccoy@nmt.edu}

\author{Alison Peck}
\affil{National Radio Astronomy Observatory, 520 Edgemont Road,
  Charlottesville, VA 22903, USA}
\email{apeck@nrao.edu}

\author{Violette Impellizzeri}
\affil{Joint ALMA Observatory, Alonso de C\'ordova 3107, Vitacura 763
  0355, Santiago de Chile, Chile}
\affil{National Radio Astronomy Observatory, 520 Edgemont Road,
  Charlottesville, VA 22903, USA}
\email{vimpelli@alma.cl}

\author{Andreas Brunthaler}
\affil{Max-Planck Institute f\"ur Radioastronomie, Auf dem H\"ugel 69,
  53121 Bonn, Germany
} 
\email{brunthal@mpifr-bonn.mpg.de}

\author{Fabian Walter}
\affil{Max-Planck Institut f\"ur Astronomie, K\"onigstuhl 17, 69117,
  Heidelberg, Germany}
\affil{National Radio Astronomy Observatory, P.O. Box O, 1003
  Lopezville Road, Socorro, NM 87801, USA}
\email{walter@mpia.de}

\author{Philip Edwards}
\affil{CSIRO Astronomy and Space Science, PO Box 76, Epping NSW 1710, Australia}
\email{Phil.Edwards@csiro.au}

\author{Crystal N. Anderson}
\affil{New Mexico Institute of Mining and Technology, 801 Leroy Place, Socorro, NM 87801, USA}
\email{canderso@nmt.edu}

\author{Christian Henkel}
\affil{Max-Planck Institute f\"ur Radioastronomie, Auf dem H\"ugel 69,
  53121 Bonn, Germany
}
\affil{Astronomy Department, Faculty of Science, King Abdulaziz University, P.O. Box 80203, Jeddah, Saudi Arabia}
\email{chenkel@mpifr-bonn.mpg.de}

\author{Ilana Feain}
\affil{CSIRO Astronomy and Space Science, PO Box 76, Epping NSW 1710, Australia}
\email{Ilana.Feain@csiro.au}

\and

\author{Minnie Y. Mao}
\affil{National Radio Astronomy Observatory, P.O. Box O, 1003
  Lopezville Road, Socorro, NM 87801, USA}
\email{mmao@nrao.edu}

\begin{abstract}
  We report the detection of a 22\,GHz water maser line in the nearest
  ($D\sim3.8$\,Mpc) radio galaxy Centaurus\,A using the Australia
  Telescope Compact Array (ATCA). The line is centered at a velocity
  of $\sim 960$\,\kms, which is redshifted by about $400$\,\kms\ from
  the systemic velocity. Such an offset, as well as the width of $\sim
  120$\,\kms, could be consistent with either a nuclear maser arising
  from an accretion disk of the central supermassive black hole, or
  for a jet maser that is emitted from the material that is shocked
  near the base of the jet in Centaurus\,A. The best spatial
  resolution of our ATCA data constrains the origin of the maser
  feature within $<3$\,pc from the supermassive black hole. The maser
  exhibits a luminosity of $\sim 1$\,L$_{\odot}$, which classifies it
  as a kilomaser, and appears to be variable on timescales of
  months. A kilomaser can also be emitted by shocked gas in star
  forming regions. Given the small projected distance from the core,
  the large offset from systemic velocity, as well as the smoothness
  of the line feature, we conclude that a jet maser line emitted by
  shocked gas around the base of the AGN is the most likely
  explanation. For this scenario we can infer a minimum density of the
  radio jet of $\gtrsim 10$\,\vden, which indicates substantial mass
  entrainment of surrounding gas into the propagating jet material.
\end{abstract}


\object{Centaurus A}

\section{Introduction}
\label{sec:intro}
Centaurus\,A (\cena) is by far the most nearby radio
galaxy and one of the brightest radio sources in the sky. Its
proximity of only $3.8$\,Mpc \citep{har10} makes it a unique target
for studies of supermassive black hole (SMBH) accretion, jet formation
and acceleration, as well as the interaction of the jets and lobes
with the interstellar and intergalactic media.

Centaurus\,A \citep[see][for a review of the properties of
Cen\,A]{isr98} is embedded in the giant elliptical host
NGC\,5128. Unlike most other elliptical galaxies, NGC\,5128 displays
very prominent dust lanes that are likely due to merger activity in
the past \citep[see e.g.][and references
therein]{isr98,bar02,aul12}. NGC\,5128 may thus be in the process of
rebuilding a strongly warped, gaseous disk
\citep[e.g.][]{qui06,neu07,str10,qui10}, which shows signatures of
spiral arms \citep{esp12}. These properties may affect the gas
accretion process onto the active SMBH (mass: $\sim5.5\times
10^{7}$\,\msun; \citealt{cap09}).

22\,GHz water (\hto) masers have proven to be great tools to study the
environment close to SMBHs. Water masers are collisionally pumped
\citep[e.g.][]{kyl91} and the requisite high gas densities are
typically achieved in shock fronts. Thus, \hto\ masers can be found in
very diverse environments. Typical locations include shells of
expanding asymptotic giant branch stars, expanding envelopes of young
stellar objects, accretion disks of SMBHs and interaction zones of
radio jets with the surrounding media \citep[see][for a
review]{lo05}. The latter two are often referred to as `disk masers'
\citep[e.g.][]{gre95,lo05} and `jet masers'
\citep[e.g.][]{cla98,pec03}. Studying nearby jet and disk masers may
also lead to a better understanding of SMBHs in the high redshift
regime. Due to their brightness, 22\,GHz water masers have been
observed up to $z\sim2.64$ \citep{imp08,cas11}.

In the past, \cena\ was target of a number of unsuccessful \hto\
maser emission searches \citep[e.g.][]{cla86,bra96,bra04,sur09}. 
In the following, we report the first detection of a water maser
toward the nuclear environment of Cen\,A. Our observations and data
reduction steps are described in \S\,\ref{sec:obs}, followed by a
presentation of our results in \S\,\ref{sec:res}, and a discussion on
the nature of the detected maser feature in \S\ref{sec:dis}. We
summarize our findings in \S\ref{sec:sum}.

\section{Observations and Data Reduction}
\label{sec:obs}
We observed Cen\,A with the Australia Telescope Compact Array
(ATCA). The ATCA is a 6 antenna interferometer with baselines between
30\,m and 6\,km. The observations were performed around the rest
frequency of the $J=6_{16}-5_{23}$ water transition at 22.23508
\,GHz. To exclude cross-identification of instrumental features with
the water line, we slightly shifted the central frequencies between
some of the observations (see Table\,\ref{tab:obs}). We used the
Compact Array Broadband Backend \citep[CABB;][]{wil11} in two
different modes: A wideband mode, with a total bandwidth of two times
1\,GHz and a channel width of 1\,MHz (or 13\,\kms\ velocity
resolution; CFB\,1M-0.5k mode) and CFB\,64M-32k, where we tuned one of
the 64\,MHz wide zooms to water (total velocity range of $\sim
860$\,\kms; channel width of 23\,kHz or $0.4$\,\kms). To reduce
systematic effects, we observed at different central frequencies (see
Table\,\ref{tab:obs}). After cropping bandpass edges, the velocity
  range of the narrow band observations amounts to $\sim
  620-1380$\,\kms\ for a central frequency of 22.172\,GHz and
  $\sim440-1290$\,\kms\ for 22.272\,GHz. Absolute fluxes were calibrated with the ATCA
  standard PKS\,1934-638. Bandpasses were typically obtained by
  observing PKS\,1253-055 for $\sim 15$\,min, but we used
  PKS\,0537-441 when the former was not available at the start of the
  observations. The observational setups and properties are summarized
  in Table\,\ref{tab:obs}, where we list the dates of the observations
  in column (1), the ATCA array configurations in column (2), the
  central frequencies and bandwidths of the instrumental setups in
  columns (3) and (4), the beam sizes and position angles in columns
  (5) and (6), the continuum fluxes in column (7), and the rms noise
  values, the channel separations, and the on-source integration times
  in columns (8), (9), and (10), respectively.

The data were self-calibrated on the strong continuum emission from
the core of \cena\ to correct phase and gain variations as a function
of time. Finally, we extracted spectra from the images, after robust
weighting and cleaning 1000 iterations per channel. All data reduction
steps were performed in MIRIAD.
Whereas the narrowband
observations show sufficiently flat baselines, fitted well by 0th order
polynomials, the wideband observations exhibit variations of order
$\sim1-2$ per cent. Given the relatively high flux density of the
central source (see Table\,\ref{tab:obs}) these variations amount to
a $\sim 100$\,mJy level, and we fitted and subtracted third order
polynomials to a $\sim 2000$\,\kms\ range around the water frequency
for satisfactory results. Given the spectral variations of the
baselines, we estimate $\sim 10$\,per cent uncertainty for the
absolute fluxes of the well behaved narrowband observations, and a
conservative $\sim 15$\, per cent error for the wideband data.

\section{Results}
\label{sec:res}

\begin{figure}
\plotone{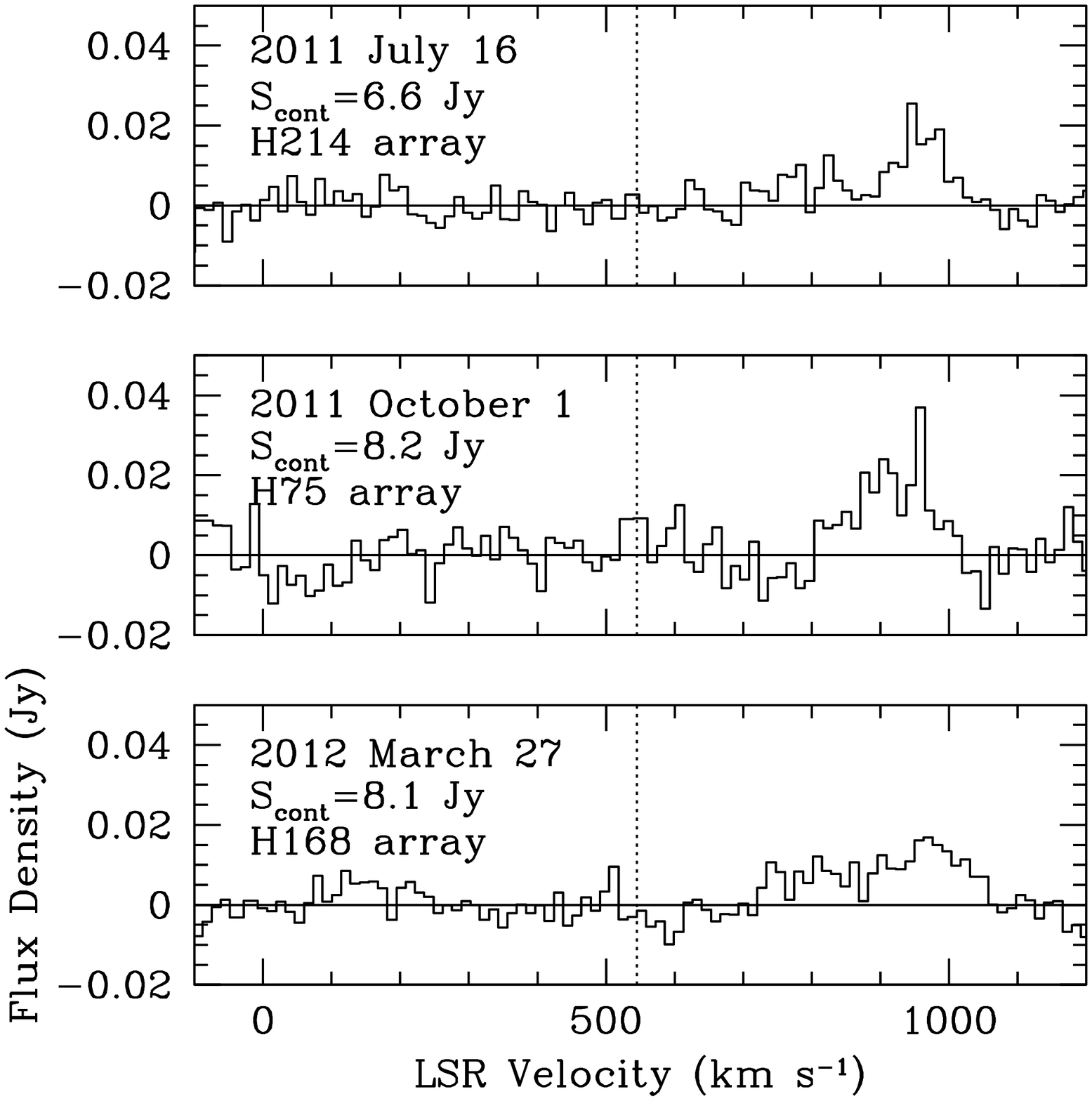}
\caption{Spectra from three different epochs of ATCA observations in
  the broadband mode. The dashed vertical lines indicate the systemic
  velocity of NGC\,5128 (545\,\kms\ LSRK). The shown spectra cover the
  $-100$\,\kms\ to $1200$\,\kms\ range of the full 1\,GHz
  observations.  \label{fig:specbroad}}
\end{figure}

\begin{figure}
\plotone{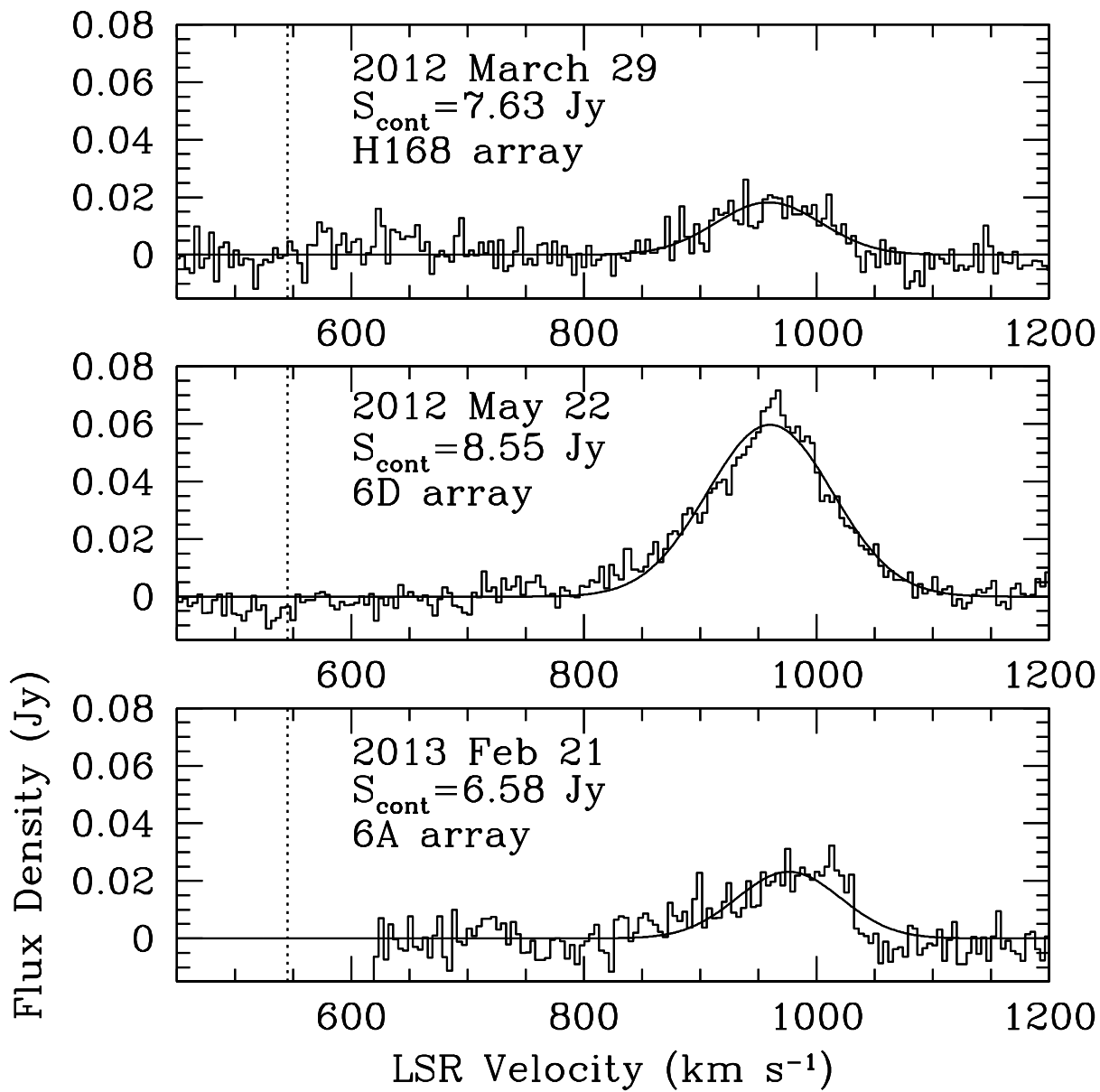}
\caption{Narrowband spectra of the \cena\ water maser line at three
  different epochs. The dashed vertical lines again indicate the systemic
  velocity of NGC\,5128 (545\,\kms\ LSRK). The best fitting Gaussian is
  shown on top of each spectrum. \label{fig:specnarrow}}
\end{figure}

We extracted line emission at the central position of the SMBH at
RA(J2000)=13$^h$25$^m$27$\fs$6,
DEC(J2000)=$-$43$^\circ$01$\arcmin$09$\arcsec$ and subtracted the
continuum emission. The wideband spectra are shown in
Fig.\,\ref{fig:specbroad} and the narrowband data in
Fig.\,\ref{fig:specnarrow}. The systemic velocity of Cen\,A is at a
kinematic Local Standard of Rest (LSRK) value of $\sim 545$\,\kms\
\citep{sav07} and we mark this value in the figures, assuming that the
line can be identified as the $J=6_{16}-5_{23}$ water maser transition
at a rest frequency of 22.23508\,GHz (see \S\ref{sec:id}).  We clearly
detect a spatially unresolved emission feature at $\sim 950$\kms\
offset from systemic.  No detections are seen at systemic and
blueshifted velocities. As the narrowband observations resulted in
higher quality spectra, we used these data to constrain the line
properties. Although the lines do not appear to be perfectly Gaussian
in shape, we nevertheless fitted one-component Gaussians to these
lines. The results are listed in Table\,\ref{tab:gauss} and plotted in
Fig.\,\ref{fig:specnarrow}. The table contains the dates of the
observations in column (1), followed by the best fit central LSRK
velocities $v_{c}$, the full widths at half maximum of the fitted line
$v_{\rm FWHM}$ and the peak fluxes $S_{p}$ in columns (2), (3), and
(4). In column (5) we list the derived isotropic luminosity values in
$L_{\odot}$.  The resulting parameters reveal that the \hto\ maser
line is at a velocity of $\sim 960$\,\kms\ redshifted $\sim
415$\,\kms\ from systemic, with a width of $\sim 120$\,\kms. The
amplitude is $\sim 20$\,mJy with the exception of the data taken on
2012 May 22, where we find a three times higher peak flux density of
$\sim 60$\,mJy.  The continuum flux at this epoch is only about 30 per
cent higher than the lowest measured continuum flux at all epochs,
whereas the line is $\sim3$ times stronger. Although some continuum
variations could be due to the different synthesized beam sizes of the
individual observations (see Table\,\ref{tab:obs}), this suggests that
the variation is predominantly intrinsic to \cena.

\section{Discussion}
\label{sec:dis}

\subsection{Location of the Line Emission}
\label{sec:loc}
Our observations are consistent with unresolved line emission emerging from
the center of \cena. For each of the 6\,km array data, the beams (see
Table\,\ref{tab:obs}) correspond to physical resolutions of about
$14.6$\,pc $\times5.9$\,pc and $22.7$\,pc $\times6.6$\,pc. The
position angles of the two observations are separated by $\sim
40^\circ$, and we can assume that the line is emitted from within the
overlap region with a conservative upper limit of a maximum of
$<3$\,pc projected distance away from the SMBH.

\subsection{Line Identification}
\label{sec:id}
We observe the line feature at a sky frequency of $\sim
22.162$\,MHz. If the line would originate in Cen\,A at its systemic
velocity of 545\,\kms\ LSRK, the line rest frequency would be expected
to be close to a value of $\sim 22.202$\,MHz. Adding a potential
velocity offset of $\pm 1000$\,\kms\ (or $\pm 75$\,MHz), the following
astrophysical spectral lines other than the \hto\ $J=6_{16}-5_{23}$
maser transition with a rest frequency of $22.23508$\,GHz may be
candidates for the emission in Cen\,A\footnote{as taken from the Lovas
  catalogue:
  \url{http://physics.nist.gov/cgi-bin/micro/table5/start.pl}}:
CH$_2$OHCHO $J=4_{13}-4_{04}$ at a rest frequency of 22.143\,MHz,
C$^{13}$CS $J=1_2-0_1$ at $\sim 22.255$\,GHz, and CCO $J=1_2-0_1$ at
22.258\,GHz. None of these lines are expected to be bright enough to
explain the observed intensity. And none of them lie near systemic
velocity requiring very unusual kinematics if they are within a star
forming cloud that is at a projected distance no further than 3\,pc
from the core of Cen\,A (\S\,\ref{sec:loc}). In addition, given the
strong continuum of Cen\,A, all of the thermal lines are more likely
to be detectable in absorption and not emission \citep[\cena\ has a
complex molecular absorption spectrum as, e.g., shown
in][]{eck90,isr92,wik97,isr98,mul09,esp10}.
Apart from the molecules, the H\,83$\beta$
radio recombination line has a rest frequency of 22.196\,GHz and could be a
potential candidate for the emission. But given that we do not even
see any H$\alpha$ recombination lines in our spectra, we can exclude
an identification with the much weaker H$\beta$ line.  The only likely
line candidate is the redshifted \hto\ maser $J=6_{16}-5_{23}$ transition,
which is by far the brightest astrophysical transition across the
observed frequency range. As alluded to in the introduction
(\S\,\ref{sec:intro}), \hto\ has also been observed toward the nuclei
of other active galaxies, and the maser line is the most likely one
observed in emission in the presence of Cen\,A's strong continuum
emission.

\subsection{Maser strength}
\label{sec:str}
The maser feature has a FWHM of $\sim 120$\,\kms, which we cannot
decompose into more narrow individual maser features down to our
$\sim1$\,\kms\ velocity resolution. We may thus assume that the line
emerges from a single maser spot unless high resolution imaging will
break it up into more components. The isotropic luminosity of the
feature is $\sim 1$\,$L_{\odot}$ (see Table\,\ref{tab:gauss}). If the
maser line is emitted from a single spot, it falls into the category of
kilomasers. \hto\ masers of similar strengths are found in nearby,
moderately star forming galaxies and mergers such as M\,51, NGC\,253,
M\,82, NGC2146, NGC\,3556, NGC\,3620 or the Antennae
\citep[][]{cla84,ho87,nak88,tar02,hen04,lo05,hen05,sur09,bru09,bro10}. It
is also similar to the Galaxy's most luminous water maser in W49N
\citep[e.g.][]{wal82}.

In all but one epoch the peak of the maser feature is about
20\,mJy. The 2012 May 22 data, however, reveal a three times stronger
line (see \S\,\ref{sec:res}). The variability may be related to a
merely $\sim 30$\,per cent stronger continuum flux at the time, which may
indicate that the maser itself is unsaturated. To determine this
conclusively, however, requires more epochs of observations,
desirably at higher spatial resolution.

\subsection{Star Formation, Jet, or Disk Maser?}

Most kilomasers are found in star formation regions of galaxies
\citep[see references above and, e.g.][]{bau96}. They typically arise
from outflows of massive stars and show velocity spreads of up to a
few tens of \kms. For Cen\,A, we can pinpoint the maser to within
$<3$\,pc projected distance from the nucleus (\S\,\ref{sec:loc}). We
consider the probability of crossing a massive star formation region
along the line of sight to be low. Even if such a region is located
toward that direction, the velocity offset of $\sim 400$\,\kms\ from
systemic and the linewidth of $\sim 120$\,\kms\, would be inconsistent
with a typical \hto\ maser in a star forming region well outside
the nuclear region.

Strong water masers are also observed toward the accretion disks of
some SMBHs (``disk masers'') and close to AGN jet-ISM interaction
regions (``jet masers''). Typical examples for disk masers are the
archetypical NGC\,4258 galaxy \citep[e.g.][]{gre95,lo05} and a larger
sample of such masers have been observed in dedicated search campaigns
\citep[e.g.][]{tar02,hen05,kon06,bra08,cas08,ben09,sur09,kon12}. Disk
masers typically have three main components: one at systemic velocity
as well as blueshifted and redshifted line features, both offset by a
few hundred \kms. Some disks, however, lack one or two of these main
components. At high velocity resolution \citep[see, e.g.][]{bra08} the
disk maser components show significant substructures at \kms\
linewidth scales. They can overlap and blend to form broader
components. Although we may see some indications of substructure in
the wideband data that was observed in 2011 (see
Fig.\,\ref{fig:specbroad}), the overall \hto\ spectrum in \cena\ is
relatively smooth, even at the \kms\ scales of our high spectral resolution data
(Fig.\,\ref{fig:specnarrow}). We thus consider a disk maser origin
unlikely. 

Jet masers, have been detected in NGC\,1068 \citep{gal96}, NGC\,3079
\citep{tro98}, Circinus \citep{gre03}, Mrk\,348 \citep{pec03} and
possibly in NGC\,1052 \citep[][]{cla98,kam05}. In particular, the jet
maser in Mrk\,348 resembles Cen\,A's maser in many respects. The line
is $\sim 130$\,\kms\ redshifted from systemic, 130\,\kms\ wide and
relatively smooth. Very Long Baseline Array data shows that the maser
is about $\sim1$\,pc away from the core. Monitoring also shows some
significant variability on timescales as short as a day and loosely
correlated with the continuum strength \citep{pec03}. The isotropic
luminosity of the Mrk\,348 maser, however, is about two orders of
magnitude higher than the feature in Cen\,A. We also observe
similarities with the possible jet maser in NGC\,1052. The \hto\ line 
in NGC\,1052 is redshifted $\sim 400$\,\kms\ from systemic with a
linewidth of $\sim 200$\,\kms. Its isotropic luminosity is of the same
order as the maser in Mrk\,348. Given the similarities of the \cena\
\hto\ maser emission with the above cases, we discuss the
implications of the jet maser model for \cena\ in the following section.

\subsection{Jet Maser Properties}
A useful model for a jet maser has been established by
\citet{pec03}. They propose that the jet plows into a neighboring
molecular cloud and creates a shock front that is propagating away
from the jet direction. The tip of the shock front covers a large
opening angle, and the morphology resembles that of a mushroom, with
cloud material being compressed toward the front and the the sides of
the jet. The projected velocities of the shocked material may thus
combine to the large linewidth that we witness. Shocks propagate
according to the density contrast between the the two media at the
shock boundary via $\rho_i v_i^2 =\rho_o v_o^2$, where $\rho_i$ and
$v_i$ are the density and velocity of the initial, shocking material
and $\rho_o$ and $v_o$ of the propagating forward shock front in the
material that was hit.  \citet{tin98} monitored the motions of the
parsec-scale jet components in Cen\,A and derived velocities of
$v_i\sim 0.1$\,c. For a shock that propagates with the velocity offset
of the maser feature in Cen\,A, $v_o\sim400$\,\kms, we can then derive
a density contrast of $\sim 2\times10^{-4}$. Interstellar \hto\ masers
require volume densities of at least $>10^{7}$\,\vden\
\citep{yat97,lo05} to be pumped \citep[see][for scenarios when
radiative pumping may become important]{kyl91}. When we apply the
density contrast to this number we can infer an incoming jet density
of $>10^{3}$\,\vden. Geometry constraints will change the numbers but
the order of magnitude estimate should remain if the shock front has a
large opening angle as suggested by \citet{pec03}.  The jet velocity,
however, may be underestimated. \citet{tin98} shows that the northern
jet is much brighter than the southern counterpart, which can be
explained by Doppler boosting the brightness of the approaching
jet. Such an effect would be expected at much higher velocities than
the $\sim 0.1c$ measured for the individual components. For the limit
of a jet velocity at light speed the above calculations will reduce
the lower limit of the jet density by two orders of magnitude and we
thus estimate a more conservative $\rho_{i}>10$\,\vden. But even such
a reduced value is much larger than typical electron-positron or
thermal gas plasma densities expected for relativistic jets. The radio
galaxy 3C120, for example, has densities $\lesssim 10^{-3}$\,\vden\
\citep{wal87}, which is four orders of magnitude lower than our limit
for the jet of \cena. The difference can potentially be explained by
surrounding material that is evaporated by and entrained into the jet
during its outward propagation.

\citet{kam05} find indications for the appearance of narrow components
in the otherwise broad maser spectrum of NGC\,1052. These components
show flux variabilities and line narrowing on timescales of days. Our
2011 ATCA broadband \hto\ spectra of \cena\ show indications for
similar, transient narrow features. In the jet maser model such
variations could be due to the jet shocking small, high density
substructures of the molecular cloud. The maser properties could be
the response to the shock dynamics that includes local density
enhancements competing with \hto\ shock dissociation.

The \cena\ maser line is redshifted to velocities larger than
systemic. The receding, fainter jet may thus be a natural location for
the origin of the jet maser feature. We note, however, that Very Long
Baseline Array data show that the similarly redshifted \hto\ maser in
Mrk\,348 originates from the approaching, brighter jet component of
this galaxy. The full location and geometry of the maser spots in
\cena\ are therefore uncertain until higher resolution imaging will be
available.

\section{Summary}
\label{sec:sum}

We report the detection of a 22\,GHz $J=6_{16}-5_{23}$ water maser
emission line toward the inner 3\,pc of the most nearby radio galaxy
Centaurus\,A. The weak and slightly variable maser feature can be
classified as a kilomaser and could have its origin in a star
formation region, a nuclear disk around the central supermassive black
hole, or from material that is shocked close to the jet base of
Cen\,A. Given the relative smoothness and width of the line, as well as its
central location, we conclude that the ``jet maser'' scenario is the
most likely one. In this case, the radio jet is shocking a central
molecular cloud. From the shock properties, a lower limit to the
density of the jet of $>10$\,\vden\ can be derived. A full
characterization of the nature and properties of the maser will
require higher spatial resolution. The discovery of a nuclear water
maser in \cena\ opens up further opportunities to probe the extreme
environments very close to a SMBH.

\acknowledgments 

We thank the Australia Telescope National Facility staff for their
assistance, and gratefully acknowledge the allocation of some
Director's Time to confirm this feature. This research has made use of
NASA's Astrophysics Data System and the NASA/IPAC Extragalactic
Database (NED) which is operated by the Jet Propulsion Laboratory,
California Institute of Technology, under contract with the National
Aeronautics and Space Administration.

{\it Facilities:} \facility{ATCA}

\clearpage

\newpage

\clearpage

\begin{deluxetable}{llcccccccc}
\tabletypesize{\scriptsize}
\tablecaption{Summary of the ATCA observations. \label{tab:obs}}
\tablewidth{0pt}
\tablehead{
\colhead{Date} & \colhead{Array} & \colhead{$\nu_{center}$} &
\colhead{BW} & \colhead{beam size}  & \colhead{Beam PA} & \colhead{$S_{\rm
    cont}$} & \colhead{rms} & \colhead{$\Delta v$} & \colhead{$t_{int}$}\\
\colhead{} & \colhead{} & \colhead{(GHz)} & \colhead{(MHz)} &
\colhead{($\arcsec$)} & \colhead{(deg)} & \colhead{(Jy)} &
\colhead{(mJy\,beam$^{-1}$)} & \colhead{(\kms)} & \colhead{(h)}\\
\colhead{(1)}&\colhead{(2)}&\colhead{(3)}&\colhead{(4)}&\colhead{(5)}&\colhead{(6)}&\colhead{(7)}&\colhead{(8)}&\colhead{(9)}&\colhead{(10)}}
\startdata
2011 July 16 & H\,214 & 22.450 & 1000 & 8.32$\times$6.59 & 82.3 & $6.6\pm1.0$ & 10 & 13 & 8 \\
2011 October 1 & H\,75 & 22.450 & 1000 & 26.80$\times$20.68 & 87.9 & $8.2\pm1.2$ & 5 & 13 & 9\\
2012 March 27 & H\,168 & 22.494 & 1000 & 12.22$\times$8.84 & 82.7 & $8.1\pm1.2$ & 4 & 13 & 8 \\
\tableline
2012 March 29 &  H\,168 & 22.172 & 64 & 12.30$\times$9.46 & 77.5 & $7.63\pm0.76$ & 3 &
0.4 & 10 \\
2012 May 22 & 6\,D & 22.172 & 64 & 0.79$\times$0.32 & 18 &  $8.66\pm0.87$  & 6 & 0.4
& 8 \\
2013 February 21 & 6\,A & 22.162 & 64 & 1.23$\times$0.36 & -26.5 & $6.58\pm0.66$ &
15 & 0.4 & 5.5
\enddata
\end{deluxetable}

\begin{deluxetable}{lcccc}
\tabletypesize{\scriptsize}
\tablecaption{Properties of the Gaussian fits to the narrowband spectra. \label{tab:gauss}}
\tablewidth{0pt}
\tablehead{
\colhead{Date} & \colhead{$v_{\rm c}$ (LSRK)} & \colhead{$v_{\rm FWHM}$} &
\colhead{$S_{\rm p}$} & \colhead{$L_{\odot}$}\\
\colhead{} & \colhead{(\kms)} & \colhead{(\kms)} &
\colhead{(mJy)} & \\
\colhead{(1)}&\colhead{(2)}&\colhead{(3)}&\colhead{(4)}&\colhead{(5)}
}
\startdata
2012 March 29 & $959\pm4$ & $107\pm11$ & $18.2\pm2.3$ & $0.7\pm0.1$ \\ 
2012 May 22 & $960\pm1$ & $127\pm3$ & $59.8\pm6.1$ & $2.7\pm0.3$ \\
2013 Feb 21 & $965\pm5$ & $126\pm12$ & $23.1\pm2.9$ & $1.0\pm0.1$
\enddata
\end{deluxetable}

\end{document}